\documentclass[12pt,a4paper,final]{iopart}

\usepackage{iopams}  
\usepackage{graphicx}
\usepackage[breaklinks=true,colorlinks=true,linkcolor=blue,urlcolor=blue,citecolor=blue]{hyperref}

\usepackage{amsmath} 
\usepackage{amssymb,color}
\usepackage[usenames,dvipsnames]{xcolor}
\usepackage{multirow,tabularx}
\usepackage{cite}

\usepackage[french]{cleveref}
\creflabelformat{figure}{#2#1#3}

\begin{document}

\title[Towards~fibered~Raman-free~correlated~photon~sources]{Spontaneous four-wave mixing in liquid-core fibers: \\Towards~fibered~Raman-free~correlated~photon~sources}

\author{M Barbier$^{1,3}$, I Zaquine$^2$ and
P Delaye$^1$}
\address{$^1$ Laboratoire Charles Fabry, Institut d'Optique, CNRS, Univ Paris-Sud, 2 avenue Augustin Fresnel, 91127 Palaiseau, France}
\address{$^2$ Laboratoire Traitement et Communication de l'Information, CNRS, T\'el\'ecom Paristech, 46 rue Barrault, 75013 Paris, France}
\address{$^3$\textit{Current address:} Optics Laboratory, Tampere University of Technology, Korkeakoulunkatu 3, 33720 Tampere, Finland}
\eads{\mailto{margaux.barbier@institutoptique.fr}, \mailto{margaux.barbier@tut.fi}}

\begin{abstract}
We experimentally demonstrate, for the first time to our knowledge, the generation of correlated photon pairs in a liquid-core photonic crystal fiber. Moreover, we show that, thanks to the specific Raman properties of liquids, the Raman noise (which is the main limitation of the performance of silica-core fiber-based correlated photon pair sources) is highly reduced. With a demonstrated coincident-to-accidental ratio equal to 63 and a pair generation efficiency of about 10$^{-4}$ per pump pulse, this work opens the way for the development of high quality correlated photon pair sources for quantum communications.
\end{abstract}

\pacs{42.65.Lm, 42.50.Dv, 42.81.Wg}
\vspace{2pc}
\noindent{\it Keywords}: Correlated photon pairs, Spontaneous four-wave mixing, Liquid-core photonic crystal fibers

\submitto{\NJP}


\section{Introduction}
\label{section-1}

Spontaneous four-wave mixing (FWM) is a third-order nonlinear process in which two pump photons are annihilated to create a pair of signal and idler photons, whose wavelengths are governed by the energy and momentum conservation conditions \cite{BoydNLO}. The signal and idler photons from the same pair are generated simultaneously, and that is why they are said to be \emph{temporally correlated}. This process can occur inside the core of an optical fiber if the pump wavelength is close to the zero-dispersion wavelength of the fiber (to satisfy both the energy and momentum conservation conditions) \cite{AgrawalNLFO,MarhicFOPA}. Thus it is advantageous to use silica-core photonic crystal fibers (PCFs) \cite{JOSAB-30-2889}, whose dispersion properties (and particularly the zero-dispersion wavelength) can be adjusted by adequately designing the microstructuring geometry \cite{JLT-24-4729}, because it allows to choose the spectral properties (spectral emission range and spectral width) of the correlated photon pair source~\cite{OE-12-3086,OL-30-3368,NJP-8-67,OL-38-73a}.

From such a correlated photon pair source, it is possible to produce \emph{entangled photon pairs}~\cite{OL-38-73a}, which are actually the key to a lot of quantum information devices (for quantum communications, quantum cryptography, quantum computation, etc.) as well as fundamental quantum tests (violation of Bell's inequalities) \cite{QIC-1-3}. More precisely, \emph{fiber-based} correlated photon pair sources are of high interest in the field of quantum communications, because a fibered architecture allows to minimize the coupling losses when connecting the source to the other components of the telecommunication network.

In the early 2000s, the first studies about the generation of correlated photon pairs by spontaneous FWM in conventional and microstructured silica-core fibers highlighted the fact that the performance of these sources was limited in terms of quantum purity because of the spontaneous Raman scattering (SpRS) process \cite{OE-12-3737,OE-13-7832,OE-17-10290,APL-98-051101,OL-38-73a}. Indeed, the Raman spectrum of silica is very broad ($\approx 40$~THz \cite{AgrawalNLFO}) and thus, regardless of the wavelengths of the emitted correlated photons, some \emph{uncorrelated} Raman photons are also generated at these wavelengths in the silica core of the fiber. Such non-filterable noise photons strongly reduce the purity of the correlated photon pair source. Note that this issue is common to all glass fibers and waveguides (chalcogenide, fluoride...), since all amorphous materials exhibit broad Raman spectra \cite{OE-18-16206,APL-98-051101,OE-20-16807}.

One of the first approaches that have been investigated in order to overcome this limitation was to cool the device down to 77~K \cite{OL-31-1905}, or even 4~K \cite{OE-17-10290}, which results in drastically reducing the efficiency of the SpRS process. The main drawback of this method lies in the complexity of the experimental setup used for cooling, making it less practical for applications. Another approach was to exploit cross-polarized phase-matching configurations in a birefringent fiber \cite{OE-12-3737}. In such a configuration, the emitted correlated photons are cross-polarized with the pump photons, whereas the Raman photons remain mainly co-polarized with the pump photons. Consequently, the Raman photons can be separated from the correlated photons using a polarizer. However, cross-polarized Raman photons remain a problem \cite{PRA-75-023803}. By using an adequately designed microstructured fiber, it is also possible to work in the normal dispersion regime of the fiber, and thus to get a very high spectral gap between the pump photons and the generated correlated photons \cite{AgrawalNLFO,NJP-8-67}. For spectral gaps higher than 40~THz, the efficiency of the SpRS process becomes negligible. However, as shown by Rarity \textit{et al.} \cite{OE-13-534} and by Cui \textit{et al.} \cite{OL-38-5063}, multiphonon Raman scattering can still be troublesome. Finally, one can decide to replace the glass fiber by a crystalline (e.g. silicon) waveguide, with the advantage of a thin-line Raman spectrum. It is then possible to avoid the generation of Raman photons at the wavelengths of the correlated photons by adequately choosing the pump wavelength \cite{OL-31-3140,OE-14-12388,OE-16-5721,NJP-13-065005,OL-36-3413}. The drawback of this option is the high coupling losses of the waveguide to an optical fiber, which makes it ill-adapted to fibered telecommunication networks. Nevertheless silicon photonic devices have been recently widely investigated on the way towards new planar optical integrated circuits technologies \cite{OE-21-27826,OL-38-2969,OE-16-20368}.

\begin{figure}[h]
\centerline{
\includegraphics[width=11cm]{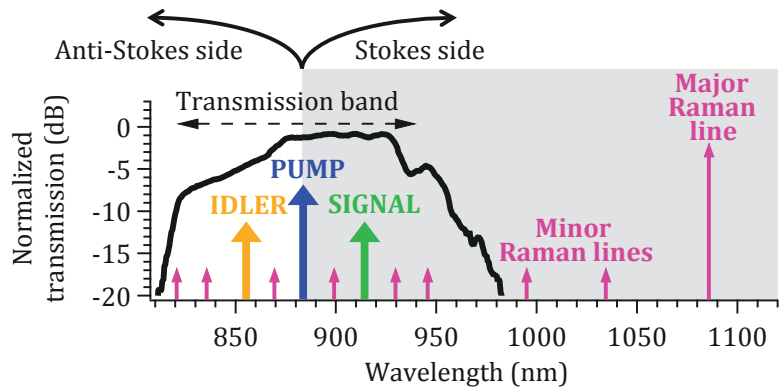}}
\vspace{-0.5\baselineskip}
\caption{Both on the Stokes ($\lambda > \lambda_p$, with $\lambda_p$ the pump wavelength) and anti-Stokes ($\lambda < \lambda_p$) sides, the major Raman lines of the liquid can be placed outside the transmission band, and the minor Raman lines that are still included can also be avoided since they are spectrally thin.$^{\rm{a}}$}
 \label{FigurePrinciple}
\vspace{0.5\baselineskip}
\scriptsize $^{\rm{a}}$ Following the convention given by Agrawal in \cite{AgrawalNLFO} and Marhic in \cite{MarhicFOPA}, the signal and idler wavelengths -- $\lambda_s$ and $\lambda_i$ respectively -- verify $\lambda_s > \lambda_p$ and $\lambda_i < \lambda_p$. \textit{Let us note that the quantum optics community usually uses the opposite denomination.}
\end{figure}

Our approach is conceptually close to this last idea, since it consists in replacing silica by a material exhibiting a thin-line Raman spectrum. In order to keep a fibered architecture and to get a compact and easy-to-use device, we have chosen to use a liquid-filled hollow-core PCF. This choice was guided by a number of previously reported investigations. First dealing with the linear properties of such liquid-core PCFs and showing that the transmission spectral band and the zero-dispersion wavelength do not only depend on the microstructuring geometry of the fiber, but also on the linear refractive index of the filling liquid \cite{OE-14-3000}. Second addressing their nonlinear properties focusing on two main axes: 1) the self-phase modulation process~\cite{AgrawalNLFO} has been exploited to measure the nonlinear refractive indices of liquids \cite{JOSAB-27-1886,JOSAB-30-1651}, and 2) the stimulated Raman scattering process has been extensively studied \cite{OE-13-4786,OL-32-337,AP-32-45,JNOPM-27-1886}, leading to the development of efficient single-mode Raman converters with high spectral flexibility. These converters take advantage of both the special Raman properties of liquids (particularly their usual thin-line Raman spectra) and the flexibility of the liquid-core fiber architecture in terms of transmission band positioning (see \cite{OE-13-4786,OL-32-337,AP-32-45,JNOPM-27-1886} for more details). \emph{Our point is that these two features can be also exploited to drastically reduce the SpRS noise in fibered sources of correlated photon pairs}: by adequately choosing the microstructuring geometry of the hollow-core PCF, the linear refractive index of the liquid, and the pump wavelength, it becomes possible to generate the photon pairs outside the \emph{thin Raman lines} of the liquid (which is not possible with the broad continua of the Raman spectra in glass materials). As a result, the Raman photons that are generated in the core of the fiber can be filtered out as their wavelengths are well-separated from those of the emitted correlated photons. Moreover, by conveniently choosing the experimental parameters, one can find an even better configuration in which \emph{the major Raman lines of the liquid (i.e. the most intense lines) are rejected outside the transmission band of the fiber} as illustrated in figure~\ref{FigurePrinciple}. In this way, the main part of the SpRS process is already filtered out by the transmission band of the fiber.

In this paper, we present the first experimental demonstration of the generation of correlated photon pairs in a liquid-core photonic crystal fiber, and we provide the experimental proof of principle for the huge potential of the liquid-core fibered architecture on the way to the development of high quality correlated photon pair fibered sources for quantum telecommunication networks. The paper is organized as follows. Section~\ref{section-2} starts with a brief description of the linear and nonlinear properties of the liquid-core fiber involved in our experiments. In section~\ref{section-3} we describe the experimental setup used to generate and detect the correlated photon pairs. Section~\ref{section-4} is devoted to our experimental results splitted in two parts: 1) detection rate measurements, providing the experimental evidence of photon pair generation from a nonlinear optics point of view, and 2) coincident count measurements, corresponding to the quantum optics vision of the correlated pairs. These results are discussed and compared to previously reported ones in section~\ref{section-5} before the conclusion is drawn in section~\ref{section-6}.

\section{Linear and nonlinear properties of the liquid-core fiber}
\label{section-2}

\begin{figure}[h]
\centerline{
\includegraphics[width=12cm]{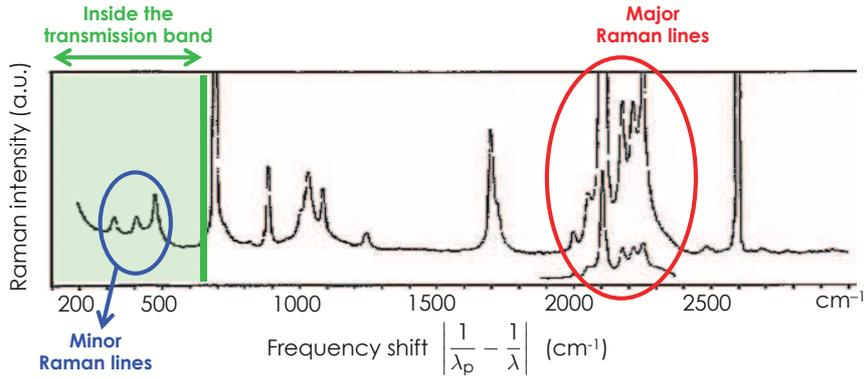}}
\vspace{-0.5\baselineskip}
 \caption{Raman spectrum of deuterated acetone (from \cite{SA-22-593}).}
 \label{FigureRaman}
\end{figure}

We use a commercially available 1-meter-long hollow-core PCF (HC-1550-PM-01 from NKT Photonics, with a core diameter of 12~$\mu$m and a cladding pitch of 4.0~$\mu$m) filled with deuterated acetone. This hollow-core PCF has been initially designed to exhibit a strong polarization-maintaining property, and once filled with deuterated acetone two principal axes can still be clearly identified (even if the birefringence is slightly reduced). The choice of the liquid relies on three main criteria: 1) its non-toxicity, 2) its negligible absorption in our working spectral range (between 700~nm and 1000~nm typically), and 3) its linear refractive index close to 1.36, which permits to shift the transmission band of the PCF from around 1550~nm (filled with air) to around 900~nm (filled with deuterated acetone), according to the shifting law given in \cite{OE-14-3000}. To experimentally confirm the value of this shift, a supercontinuum source was used to measure the transmission band of the liquid-core fiber, showing that it ranges typically from 820~nm to 925~nm (see transmission band in figure~\ref{FigurePrinciple}). As a result, the transmission band matches the spectral tunability range of the Ti:Sapphire laser used as the pump for our spontaneous FWM experiments. This transmission band measurement also confirms that the main Raman lines of deuterated acetone (with Raman shifts around 2100~cm$^{-1}$) are rejected outside the transmission band, while three minor Raman lines (at 331~cm$^{-1}$, 410~cm$^{-1}$, and 478~cm$^{-1}$) are still included (see figure~\ref{FigureRaman}) and must be avoided by adequately choosing the pump wavelength for the FWM process (this point will be discussed in more details in the following section). The 1-meter fiber length allowed a reasonable core-filling time together with a good precision on the preliminary linear and nonlinear characterization.

One-picosecond pulses emitted by a tunable mode-locked Ti:Sapphire pump laser are injected in the liquid-core fiber, with a repetition rate of 80~MHz, after filtering out the broadband residual fluorescence background of the laser to negligible level (attenuation of 50~dB at least). Even though this kind of fiber architecture, based on commercial fibers initially optimized for empty uses, is usually slightly multimode \cite{OL-32-337}, a fine adjustment allows to stably inject the pump pulses in the Gaussian fundamental mode \cite{JOSAB-27-1886} (as was easily checked in our experiments by regularly imaging the output face of the fiber on a CCD camera). The pump transmission, including both coupling and mode propagation losses, is greater than 40\% and can be as large as 50\%. Noting that some previous measurements performed with similar fibers led to propagation losses of the order of 0.75~dB.m$^{-1}$ \cite{JOSAB-27-1886}, we can reasonably consider that the greatest part of the 3-dB transmission loss  is rather related to imperfect coupling.

\begin{figure}[h]
\centerline{
\includegraphics[width=9cm]{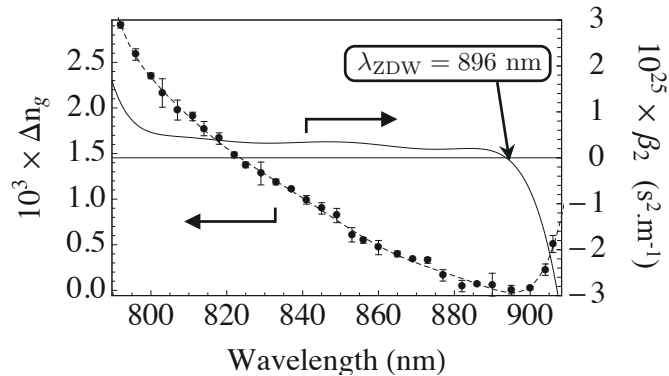}}
\vspace{-0.5\baselineskip}
 \caption{Group-velocity dispersion ($\beta_2$) curve of the liquid-core fiber (solid line), deduced from a time-of-flight measurement of the group index variations $\Delta n_g$ on the fundamental mode of the fiber (dots).}
 \label{FigureDispersion}
\end{figure}

Taking advantage of the tunability of the Ti:Sapphire pump source, the group-velocity dispersion (GVD) curve $\beta_2$ of the liquid-core fiber was experimentally determined (by an interferometric time-of-flight measurement) and led in particular to a \emph{zero-dispersion wavelength ($\lambda_{\text{ZDW}}$) equal to $(896 \pm 1)$~nm} (see figure~\ref{FigureDispersion} and Appendix~A).

Finally, by exploiting the self-phase modulation-induced spectral broadening of the pump pulses \cite{JOSAB-27-1886}, the nonlinear coefficient $\gamma$ of the fiber~\cite{AgrawalNLFO} was measured to be equal to $(0.012 \hspace{0.1em}\pm\hspace{0.1em} 0.002)$~m$^{-1}$.W$^{-1}$. Given the values of the nonlinear refractive index of deuterated acetone ($n_2 = 5.2 \times 10^{-20}$~m$^2$.W$^{-1}$ according to \cite{JOSAB-30-1651}) and of the pump wavelength ($\lambda_p \approx \lambda_{\text{ZDW}} = 896$~nm), we could deduce the effective mode area $A_{\text{eff}}$ thanks to the following formula~\cite{AgrawalNLFO}:
	$\gamma = 2 \pi n_2/(\lambda_p A_{\text{eff}})$.
We found $A_{\text{eff}} \approx 30~\mu$m$^{2}$, which is consistent with the specifications of the liquid-core fiber and with some previous measurements performed with a similar fiber (with close microstructuring geometry and same filling liquid)~\cite{JOSAB-30-1651}. These results show that the nonlinear behaviour of our liquid-core fiber is in agreement with what can be expected as far as the automatically phase-matched nonlinear process of self-phase modulation is concerned.

\begin{figure}[h]
\centerline{
\includegraphics[width=9cm]{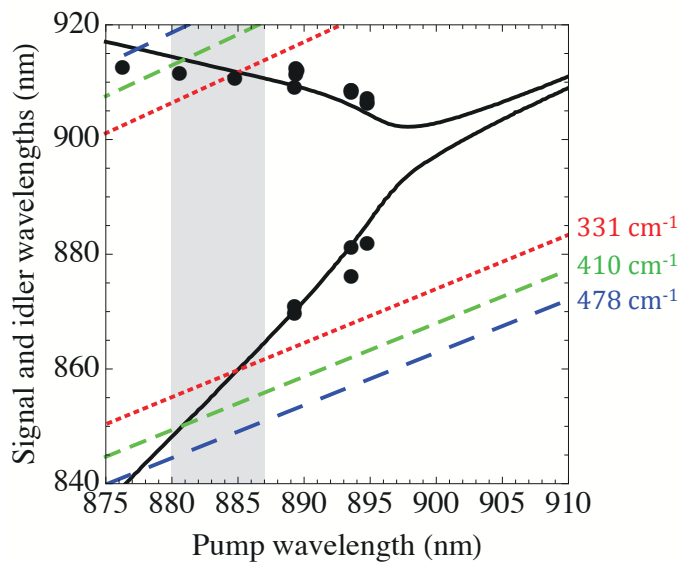}}
\vspace{-0.5\baselineskip}
 \caption{Solid line: Phase-matching curve for our liquid-core PCF, calculated from the characterization experimental data (for a peak pump power of 100~W). Dots: Signal and idler wavelengths measured in amplification regime. Dashed lines: Minor Raman lines of deuterated acetone included inside the transmission band of our liquid-core PCF. Shaded area: Working spectral range.}
 \label{FigurePMCurve}
\end{figure}

From these preliminary characterization steps, we were able to deduce the phase-matching curve for the FWM process in this fiber (see figure~\ref{FigurePMCurve}, solid line), which is given by \cite{AgrawalNLFO}:
\begin{equation}
	\fl 2 \beta(\omega_p) - \beta(\omega_s) - \beta(\omega_i) - 2 \gamma P_p = 0,
\end{equation}
with $\beta$ the propagation constant on the fundamental mode of the fiber, $\omega_p$, $\omega_s$ and $\omega_i$ the pump, signal, and idler angular frequencies respectively, and $P_p$ the peak pump power. Experiments of parametric amplification by FWM were also performed in order to confirm the validity of this phase-matching curve: in addition to the Ti:Sapphire picosecond pump pulses, we injected (in the fundamental mode of the fiber) a broadband pulsed signal \cite{JOSAB-30-2889} (temporally overlapped with the pump pulses). The output spectrum, acquired from an optical spectrum analyzer, confirmed that the amplification process actually occurs around the wavelengths predicted by the phase-matching curve (see dots in figure~\ref{FigurePMCurve}).

\section{Experimental setup}
\label{section-3}

From the phase-matching curve plotted in figure~\ref{FigurePMCurve}, we know, for a given pump wavelength, what signal and idler wavelengths will be spontaneously generated with the highest probability (note that this phase-matching curve exhibits an uncertainty of a few nanometers, on both abscissa and ordinate, due to the uncertainties on the time-of-flight GVD measurement). As a result, it allows us to define our working spectral range, which must satisfy the two following criteria:
\begin{enumerate}
	\item the spectral gap between the pump photons and the correlated signal and idler photons must be of the order of 30~nm (i.e. $\approx 11$~THz, or $\approx 375$~cm$^{-1}$), which is a compromise between a good transmission of the signal and idler photons in the fundamental mode of the liquid-core fiber (see the transmission band in figure~\ref{FigurePrinciple}) and a good filtering efficiency of the pump photons at the output of the liquid-core fiber;
	\item the minor Raman lines of deuterated acetone (dashed lines in figure~\ref{FigurePMCurve}) must be avoided.
\end{enumerate}
One can see that we have to work with a pump wavelength ($\lambda_p$) between 880~nm and 887~nm typically (shaded area in figure~\ref{FigurePMCurve}). After some optimization steps, we chose $\lambda_p = 885.5$~nm, which leads to the generation of signal and idler photons at 916~nm and 857~nm respectively.

\begin{figure}[h]
\centerline{
\includegraphics[width=12cm]{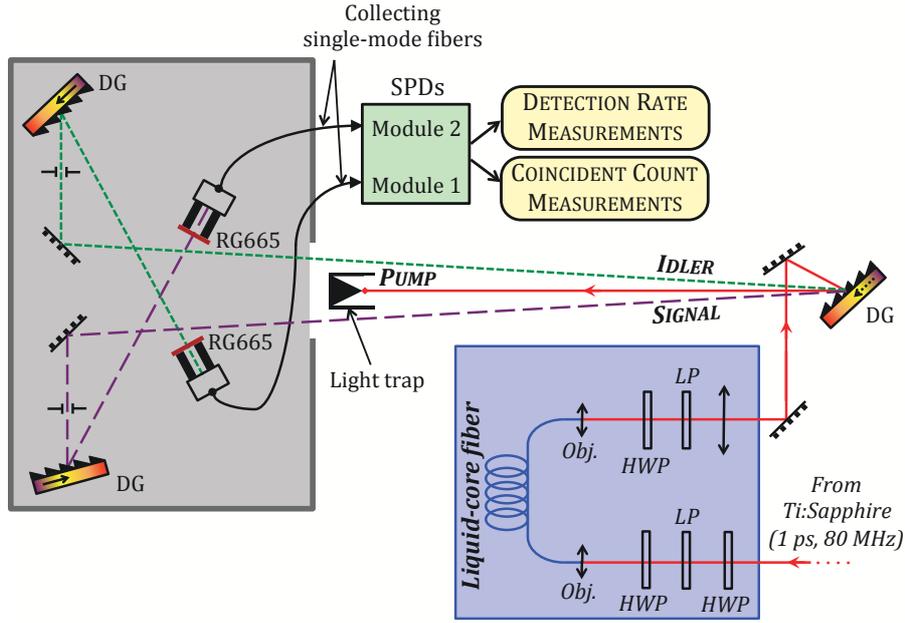}}
 \caption{Detection setup. DG = diffraction grating, SPDs = single-photon detectors, LP = linear polarizer, HWP = half-wave plate, Obj. = microscope objective, RG665 = Schott longpass filter.}
 \label{FigureSetup}
\end{figure}

In order to separate the spontaneously generated signal and idler photons from each other and from the huge amount of remaining pump photons, we built a free-space two-arm double-grating spectrometer at the output of the liquid-core fiber (see figure~\ref{FigureSetup}).

Signal (respectively idler) photons are sent into the first (resp. second) arm of the spectrometer, while the pump photons are eliminated (collected in a light trap). At the end of the arm, signal (resp. idler) photons are re-coupled into a single-mode fiber connected to a silicon avalanche-photodiode single-photon detector (SPD) (SPCM-AQRH-14 from \emph{Excelitas Technologies}, with a dark count level of about 60~s$^{-1}$, a dead time equal to 20~ns, and a detection efficiency around 40\% at 900~nm). The detection bandwidth is about 0.5 nm on both channels, and for a spectral gap between the pump photons and the correlated photons of  $\approx 30$~nm, we get a pump filtering efficiency of $\approx 130$~dB. Starting from approximately 10$^7$ photons per pulse and given the measured $\approx$~3\% collection efficiency for the correlated photons between the output of the source and the SPD (which includes the transmission of the spectrometer arm, the coupling efficiency in the collecting single-mode fiber, and the detection efficiency of the SPD), such a pump filtering efficiency allows a correct detection of the correlated photons as soon as the generation efficiency is higher than 10$^{-4}$ pairs per pulse typically (measured at the very output of the liquid-core fiber). This generation efficiency must be kept quite low, however, in order to avoid the generation of multiple pairs by the same pump pulse (which would degrade the quality of the source, as shown in \cite{JOSAB-28-832}). As a result, we have to work with generation efficiencies between 10$^{-4}$ and 10$^{-2}$ pairs per pulse typically. Given the repetition rate of the pump laser (80~MHz) and the collection efficiency ($\approx 3\%$), it means that we should keep the FWM detection rate of the two SPDs between 200 and 30~000 counts per second approximately (the \textit{raw detection rate}, which includes the FWM detection rate \emph{and} the noise detection rate -- from residual pump, Raman, environment, dark counts --, will necessarily be a little bit higher).

\section{Experimental results}
\label{section-4}

\subsection{Detection rate}
\label{subsection-4-1}

We first measured the evolution of the detection rate $\Gamma_{\text{det}}$ on each SPD as a function of the mean pump power ($P_{\text{mean}}$) injected in the fiber core. We expect the following dependence :
\begin{equation}
	\fl \Gamma_{\text{det}} = \alpha_0 + \alpha_1 P_{\text{mean}} + \alpha_2 P_{\text{mean}}^2.
\end{equation}
The constant term $\alpha_0$ originates from the dark counts of the SPD and the environmental noise. The linear term $\alpha_1 P_{\text{mean}}$ can come from the residual pump photons on the one hand, and from the Raman photons on the other hand (since the efficiency of the SpRS process grows linearly with the mean pump power). \emph{Obviously, in the case of our liquid-core fiber, we expect the SpRS contribution to be much lower than in the case of a silica-core fiber}. Finally, the quadratic term $\alpha_2 P_{\text{mean}}^2$ will be the signature of the spontaneous FWM process.

\begin{figure}[h]
\centerline{
\includegraphics[width=7.5cm]{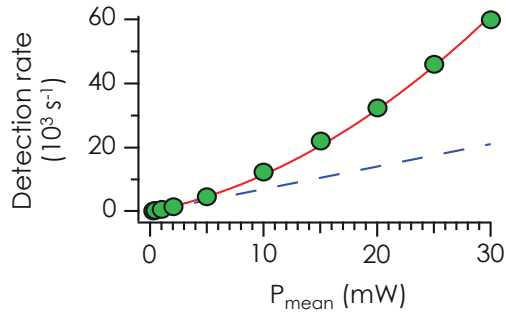}}
 \caption{Typical evolution of the detection rate as a function of the mean pump power at $\lambda_p = 885.5$~nm. Dots: Experimental data. Solid line: Second-order polynomial fit. The dashed line represents the noise contribution (i.e. $\alpha_0 + \alpha_1 P_{\text{mean}}$).}
 \label{FigureDetectionRate}
\end{figure}

A typical result of this kind of detection rate measurement is given in figure~\ref{FigureDetectionRate}. A second-order polynomial fit on the experimental data leads to $\alpha_1 = 700$~s$^{-1}$.mW$^{-1}$ and $\alpha_2 = 45$~s$^{-1}$.mW$^{-2}$ (an independent prior measurement led to $\alpha_0 \approx 60$~s$^{-1}$, showing that the environmental noise level is negligible with respect to the dark count level). We used mean pump powers between 0.2~mW and 30~mW in order to keep the FWM detection rate between 200~s$^{-1}$ and 30~000~s$^{-1}$, as explained at the end of the previous section.

Let us note that with the silica-core PCF whose characteristics are presented in the Appendix~A, we obtained $\alpha_1 \approx 9 \times 10^5$~s$^{-1}$.mW$^{-1}$ with the same pump filtering efficiency and detection setup. \emph{The value of $\alpha_1$ in the liquid-core fiber is about three orders of magnitude lower than in the silica-core PCF}. 

\begin{figure}[h]
\centerline{
\includegraphics[width=9cm]{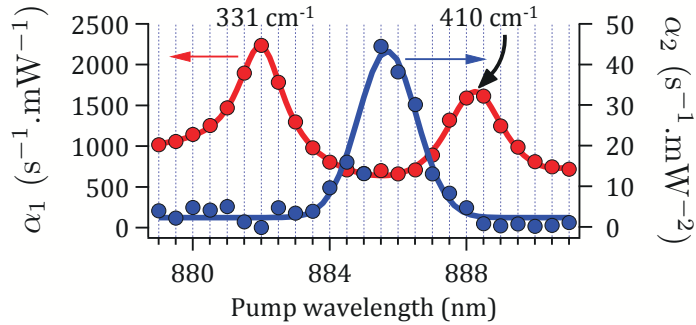}}
 \caption{Evolutions of $\alpha_1$ and $\alpha_2$ as a function of $\lambda_p$, with the idler detection arm centered at 857~nm. Solid lines: double-lorentzian fit on the $\alpha_1$ data, and Gaussian fit on the $\alpha_2$ data.}
 \label{FigureDetectionRate2}
\end{figure}

The following step was to measure the evolution of $\Gamma_{\text{det}}$ with $P_{\text{mean}}$ for different values of the pump wavelength ($\lambda_p$) inside the spectral range of interest (shaded area in figure~\ref{FigurePMCurve}), while the two arms of the spectrometer were still collecting the photons at 857~nm and 916~nm. For each $\lambda_p$, we obtained a curve such as that of figure~\ref{FigureDetectionRate}, from which we could retrieve the values of $\alpha_1$ and $\alpha_2$. Figure~\ref{FigureDetectionRate2} shows the evolution of the values of $\alpha_1$ and $\alpha_2$ as a function of $\lambda_p$. It reproduces respectively the minor Raman lines of deuterated acetone and the parametric band associated with the FWM process (convolved with the $\approx 0.5$~nm-detection bandwidth of the spectrometer). One can notice that 1) $\alpha_2$ actually reaches its maximum value when $\lambda_p$ is equal to 885.5~nm, and 2) in the chosen spectrometer configuration, the parametric band is actually located \emph{between} two minor Raman lines of deuterated acetone (the overlap between the parametric band and these Raman lines is minimized), as required. Note that this result also permits to ensure that the quadratic growth obtained on the curve of figure~\ref{FigureDetectionRate} has no chance to come from a \emph{stimulated} Raman scattering process, but actually originates from the spontaneous FWM process.

Moreover, by performing a double-lorentzian fit on the $\alpha_1$ data, we were able to estimate the values of the Raman scattering effective cross-sections of deuterated acetone for these two Raman lines. For example, considering the minor line with Raman shift equal to 331~cm$^{-1}$, our experimental data lead to an effective cross-section equal to $1.3 \times 10^{-32}$~cm$^2$.sr$^{-1}$ whereas an estimation deduced from reported data \cite{SA-22-593,JCP-56-3384,JCP-60-2556} gives $2 \times 10^{-32}$~cm$^2$.sr$^{-1}$. The good agreement between these two values confirms the validity of our measurements related to the SpRS process.

It must be noted nevertheless, that the spontaneous FWM efficiency, appears to be unexpectedly low. We actually obtained $\alpha_{2 \text{(liquid)}} = 45$~s$^{-1}$.mW$^{-2}$ (maximum of the $\alpha_2$ curve in figure~\ref{FigureDetectionRate2}), to be compared to the silica-core PCF used as a reference (see Appendix~A), for which $\alpha_{2 \text{(silica)}} \approx 48 \times 10^4$~s$^{-1}$.mW$^{-2}$. Since the nonlinear coefficient $\gamma$ of the liquid-core PCF is only $\approx 5$ times lower than that of the silica-core PCF, the FWM efficiency in our liquid-core fiber (which is proportional to $(\gamma L)^2$) is $\approx 400$ times lower than expected. It corresponds to an effective nonlinear interaction length $L$ around 20 times lower than the real length of the fiber. A temporal walk-off between the pump pulse and the signal and idler photons is often said to reduce the effective length \cite{AgrawalNLFO} but in the case of the photon pair generation in the silica-core PCF reported  in \cite{NJP-8-67}, the temporal walk-off is said to have no influence on the number of generated photons. Moreover, a simple analytical model that we have developed in order to address this issue in a more quantitative way, taking into account the pulsed temporal regime of the pump \cite{BarbierPhD}, does not lead us to attribute this low FWM efficiency to a walk-off problem. Inhomogeneities of the microstructuring structure could rather cause a variation of the zero-dispersion wavelength along the fiber, leading to a fluctuation of the phase-matching condition satisfaction, and therefore justify this reduction of the effective nonlinear interaction length. The influence of such a phenomenon would be low in the case of the silica-core PCF, as it is much less dispersive than the liquid-core fiber. This issue is currently under investigation, both experimentally and theoretically. It essentially shows that we still have a great potential of improvement on these first results, if a specially designed fiber is used for such experiments.

In summary, the results described in this section are the key to two essential claims of this paper:
\begin{itemize}
	\item  At high pump power levels the quadratic part of the growth of $\Gamma_{\text{det}}$ becomes significant, which highlights the fact that the spontaneous FWM process actually occurs in the core of our liquid-filled fiber. \emph{This is the first experimental demonstration of correlated photon pair generation in a liquid-core fiber}.
\item the SpRS contribution has been reduced by about three orders of magnitude in the liquid-core fiber compared to the silica-core one. 
\end{itemize}

\subsection{Coincident counts}
\label{subsection-4-2}

\begin{figure}[h]
\centerline{
\includegraphics[width=9cm]{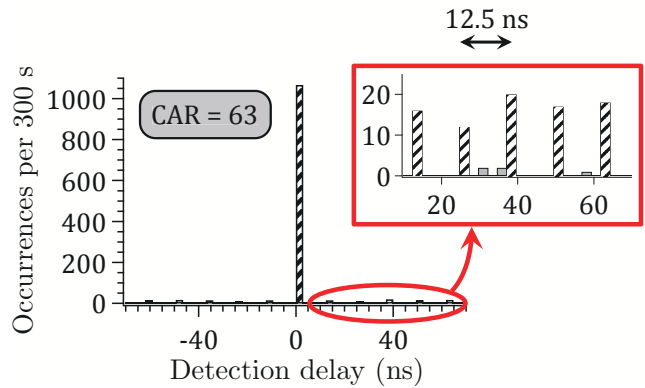}}
 \caption{Histogram of the delays between a detection on the idler channel and a detection on the signal channel, acquired with $P_{\text{mean}} =$~2.8~mW, $P_{\text{peak}} = 17.5$~W. For this measurement, the generation efficiency was of the order of $10^{-4}$ pairs per pulse, and the detection rates on the two SPDs were $\Gamma_{\text{det,s}} \approx 5850$~s$^{-1}$ and $\Gamma_{\text{det,i}} \approx 2100$~s$^{-1}$.}
 \label{FigureHisto}
\end{figure}

A second series of experiments allowed us to plot the histogram of the temporal correlations between the two SPDs, which represents the number of detection pairs (one detection on each SPD) -- called \textit{occurrences} in figure~\ref{FigureHisto} -- as a function of the delay between the two detections of the pair. In this way, we obtained the number of temporally coincident counts, whose predominance is the signature of the generation of correlated photon pairs. The coincident-to-accidental ratio (CAR), i.e. the ratio between the total number of coincidences and the number of accidental coincidences (see Appendix~B for technical details), is a quantitative criterion to estimate the performance of the whole device, including both the quantum purity of the source and the performance of the detection setup (particularly, its collection efficiency).

On the histogram of figure~\ref{FigureHisto}, we get a CAR equal to 63 for a generation efficiency of about $10^{-4}$ pairs per pulse at the output of the fiber.

\section{Discussion}
\label{section-5}

Table~\ref{table3} summarizes the parameters of interest for a valuable comparison between our experimental results and previously reported ones concerning the generation of correlated photon pairs in $\chi^{(3)}$ fibered materials: microstructured silica-core fibers \cite{NJP-8-67,OE-17-21302,OL-38-5063},  highly nonlinear non-microstructured silica fibers \cite{OL-38-73} and chalcogenide fibers \cite{OE-20-16807}.

\newlength{\La}
\setlength{\La}{0.7cm}
\newlength{\Lb}
\setlength{\Lb}{1.5cm}
\newlength{\Lc}
\setlength{\Lc}{2.1cm}
\newlength{\Ld}
\setlength{\Ld}{1.3cm}
\newlength{\Le}
\setlength{\Le}{2cm}
\newlength{\Lf}
\setlength{\Lf}{0.8cm}
\newlength{\Lg}
\setlength{\Lg}{1.5cm}
\newlength{\Lh}
\setlength{\Lh}{0.5cm}
\newlength{\Li}
\setlength{\Li}{1.7cm}

\begin{table}[h!]
  \caption{Comparison with previously reported results in fibered structures.}
  \begin{center}
  \scriptsize
    \begin{tabularx}{15.8cm}{ccccccccc}
    \br     
    \parbox{\La}{\centering Ref.} & \parbox{\Lb}{\centering Architecture} & \parbox{\Lc}{\centering Pump \\ laser$^{\rm a}$} & \parbox{\Ld}{\centering Spectral \\ detuning$^{\rm b}$} & \parbox{\Le}{\centering Detection \\ setup$^{\rm c}$} & \parbox{\Lf}{\centering T$^\circ$} & \parbox{\Lg}{\centering Mean \\ pump \\ power} & \parbox{\Lh}{\centering CAR} & \parbox{\Li}{\vspace{0.3\baselineskip}\centering Generation \\ efficiency \\ (pairs/pulse)\vspace{0.3\baselineskip}} \\
    \hline
    \hline
    \parbox{\La}{\centering\cite{NJP-8-67}\\ (2006)} & \parbox{\Lb}{\centering 0.2-m \\silica-core \\ PCF} & \parbox{\Lc}{\centering $\lambda_p = 708$~nm \\ $\Delta t \approx 2$~ps \\ $\Gamma = 80$~MHz} & \parbox{\Ld}{\centering 2940~cm$^{-1}$ \\ 88~THz \\ $\approx 150$~nm} & \parbox{\Le}{\centering \vspace{0.3\baselineskip}$\Delta\lambda_s = 4.5$~nm \\ $\Delta\lambda_i = 9.6$~nm \\ $\eta_s \approx 24$\% \\ $\eta_i \approx 11$\%\vspace{0.3\baselineskip}} & \parbox{\Lf}{\centering Room \\ T$^\circ$} & \parbox{\Lg}{\centering 340~$\mu$W} & \parbox{\Lh}{\centering 220} & \parbox{\Li}{\centering $1 \times 10^{-2}$} \\
   \hline
    \parbox{\La}{\centering \cite{OE-17-21302} \\ (2009)} & \parbox{\Lb}{\centering 1-m \\ silica-core \\ PCF} & \parbox{\Lc}{\centering $\lambda_p = 742$~nm \\ $\Delta t \approx 8$~ps \\Â $\Gamma = 76$~MHz} & \parbox{\Ld}{\centering 1000~cm$^{-1}$ \\ 30~THz \\ $\approx 55$~nm} & \parbox{\Le}{\centering \vspace{0.3\baselineskip}$\Delta\lambda_s = 0.17$~nm \\ $\Delta\lambda_i = 0.3$~nm \\ $\eta_s = 7.3$\% \\ $\eta_i = 7.7$\%\vspace{0.3\baselineskip}} & \parbox{\Lf}{\centering Room \\ T$^\circ$} & \parbox{\Lg}{\centering $\approx 0.05$~mW\vspace{0.5\baselineskip}\\ \vspace{0.5\baselineskip}0.5~mW} & \parbox{\Lh}{\centering 900\vspace{0.5\baselineskip}\\ \vspace{0.5\baselineskip}100} & \parbox{\Li}{\centering $1 \times 10^{-4}$\vspace{0.5\baselineskip} \\ \vspace{0.5\baselineskip}$0.9 \times 10^{-2}$} \\
    \hline
    \parbox{\La}{\centering \cite{OL-38-5063} \\ (2013)} & \parbox{\Lb}{\centering 0.15-m \\ silica-based \\ micro/nano-fiber} & \parbox{\Lc}{\centering $\lambda_p = 1041$~nm \\ $\Delta t \approx 250$~fs \\ $\Gamma = 62.56$~MHz} & \parbox{\Ld}{\centering 2000~cm$^{-1}$ \\ 60~THz \\ $\approx 230$~nm} & \parbox{\Le}{\centering \vspace{0.3\baselineskip}$\Delta\lambda_s = 18$~nm \\ $\Delta\lambda_i = 9$~nm \\ $\eta_s \approx 2$\% \\ $\eta_i \approx 10$\%\vspace{0.3\baselineskip}} & \parbox{\Lf}{\centering Room \\ T$^\circ$} & \parbox{\Lg}{\centering 1~mW} & \parbox{\Lh}{\centering 530} & \parbox{\Li}{\centering $\approx 1 \times 10^{-3}$} \\
 \hline
     \hline
    \parbox{\La}{\centering \cite{OL-38-73} \\ (2013)} & \parbox{\Lb}{\centering 10-m \\ silica-based \\ HNLF} & \parbox{\Lc}{\centering $\lambda_p = 1554.1$~nm \\ $\Delta t \approx 5$~ps \\Â $\Gamma = 46.5$~MHz} & \parbox{\Ld}{\centering 26.7~cm$^{-1}$ \\ 0.8~THz \\ 6.5~nm} & \parbox{\Le}{\centering \vspace{0.3\baselineskip}$\Delta\lambda_s \approx 1$~nm \\ $\Delta\lambda_i \approx 1$~nm \\ $\eta_s = 4.99$\% \\ $\eta_i = 4.87$\%\vspace{0.3\baselineskip}} & \parbox{\Lf}{\centering 300 K\vspace{0.5\baselineskip}\\ \vspace{0.5\baselineskip}77 K} & \parbox{\Lg}{\centering $\approx 430~\mu$W} & \parbox{\Lh}{\centering 29\vspace{0.5\baselineskip}\\ \vspace{0.5\baselineskip}130} & \parbox{\Li}{\centering $1 \times 10^{-4}$\vspace{0.5\baselineskip} \\ \vspace{0.5\baselineskip}$7 \times 10^{-5}$} \\   
   \hline
    \hline
 \parbox{\La}{\centering \cite{OE-20-16807} \\ (2012)} & \parbox{\Lb}{\centering 0.07-m \\ chalcogenide \\ fiber} & \parbox{\Lc}{\centering $\lambda_p = 1550.1$~nm \\ $\Delta t \approx 10$~ps \\Â $\Gamma = 10$~MHz} & \parbox{\Ld}{\centering 23~cm$^{-1}$ \\ 0.7~THz \\ 5.6~nm} & \parbox{\Le}{\centering \vspace{0.3\baselineskip}$\Delta\lambda_s \approx 0.4$~nm \\ $\Delta\lambda_i \approx 0.4$~nm \\ $\eta_s = 3.0$\% \\ $\eta_i = 2.2$\%\vspace{0.3\baselineskip}} & \parbox{\Lf}{\centering 293 K\vspace{0.5\baselineskip}\\ \vspace{0.5\baselineskip}77 K} & \parbox{\Lg}{\centering ///} & \parbox{\Lh}{\centering 0.5\vspace{0.5\baselineskip}\\ \vspace{0.5\baselineskip}4.2} & \parbox{\Li}{\centering $\approx 4 \times 10^{-3}$} \\
    \hline
 \hline
\parbox{\La}{\centering This \\ paper} & \parbox{\Lb}{\centering 1.05-m \\ liquid-core \\ PCF} & \parbox{\Lc}{$\centering \lambda_p = 885.5$~nm \\ $\Delta t \approx 1$~ps \\Â $\Gamma = 80$~MHz} & \parbox{\Ld}{\centering 375~cm$^{-1}$ \\ 11.3~THz \\ 30~nm} & \parbox{\Le}{\centering \vspace{0.3\baselineskip}$\Delta\lambda_s \approx 0.5$~nm \\ $\Delta\lambda_i \approx 0.5$~nm \\ $\eta_s \approx 3$\% \\ $\eta_i \approx 3$\%\vspace{0.3\baselineskip}} & \parbox{\Lf}{\centering Room \\ T$^\circ$} & \parbox{\Lg}{\centering 2.8~mW} & \parbox{\Lh}{\centering 63} & \parbox{\Li}{\centering $1 \times 10^{-4}$} \\
     \br   
    \end{tabularx}
  \end{center}
  \label{table3}
  \scriptsize
$^{\rm a}$ $\lambda_p$ = pump wavelength, $\Delta t$ = pulse duration, $\Gamma$ = repetition rate.

$^{\rm b}$ Warning: For detunings higher than 10~THz typically, the value given in terms of wavelength is only a rough value!

$^{\rm c}$ $\Delta\lambda_{s,i}$ = Detection bandwidth in the signal (resp. idler) channel, $\eta_{s,i}$ = collection efficiency in the signal (resp. idler) channel. "Signal" and "idler" follow the convention taken in the related article (which can differ from our convention).
\end{table}

\normalsize

This comparison is complex, because of the numerous parameters involved in the interpretation of the obtained performances. It is noteworthy that the CAR is expected to decrease with increasing generation efficiency and that this effect is all the more important as the losses experienced by the signal and idler photons are high, hence the importance of the collection efficiency of the photon pairs. The generation rate is expected to increase with the detection bandwidth of the signal and idler photons. The best reported performances \cite{NJP-8-67,OE-17-21302,OL-38-5063} combining a high CAR and a high generation efficiency correspond either to very high spectral detuning which reduces the noise contribution (at least from SpRS) or to optimized coupling efficiency of the signal and idler photons (or both). Compared to this, the performance of this first demonstration in a commercial fiber, with a relatively small detuning and one of the smallest detection bandwidths, is very promising. We already achieve a CAR value better than what is reported in the literature in the case of room temperature silica-core fiber sources of correlated photon pairs with low detuning between the pump and correlated photons: in \cite{OL-38-73}, Sua \textit{et al.} use a 10-m-long silica-based highly nonlinear fiber (HNLF) and report a CAR equal to 29 at room temperature (they achieve a CAR of 130 by cooling the fiber down to 77~K), with a generation efficiency of the same order as ours, around 10$^{-4}$.

In contrast with previously reported results, we believe that the main limitation of our device \emph{is not the SpRS process anymore}: indeed, as shown in the previous subsection, the SpRS contribution has been drastically reduced by replacing the silica core by a liquid core.

\section{Conclusion}
\label{section-6}

In this paper, we experimentally demonstrate, for the first time to our knowledge, the generation of correlated photon pairs in a liquid-filled hollow-core photonic crystal fiber (PCF). Moreover, we provide the proof of principle for the high potential of this hybrid architecture on the issue of reducing the spontaneous Raman scattering (SpRS) noise in fiber-based correlated photon pair sources. Indeed, by comparing the experimental results obtained in a liquid-core PCF and those obtained with a silica-core PCF, we evidenced a reduction of the SpRS contribution by about three orders of magnitude.

We demonstrate a coincident-to-accidental ratio (CAR) of 63 with a generation efficiency of about $10^{-4}$~pairs per pulse obtained with a mean pump power of 2.8~mW, a peak pump power of 17.5~W, and detection bandwidth of about 0.5~nm for the signal and idler photons, and we expect to be able to highly improve these performances. We will focus on improving the spontaneous four-wave mixing efficiency, by optimizing the dispersive and nonlinear properties of the hollow-core PCF. A design specifically adapted to the targeted application together with the use of various nonlinear liquids will increase the pair generation rate and further minimize the possible residual SpRS contribution of the silica structure of the fiber and of the liquid (the latter could originate from a slight overlap between the minor Raman lines of deuterated acetone, or from the leakage of some low-frequencey Raman lines \cite{JPCB-105-6004,JOSAB-29-1977}). The choice of the fiber length will also be of significant importance. All these issues are currently under investigation, both experimentally and theoretically, in order to move forward on the road to the development of high quantum quality fiber-based photon pair sources for quantum telecommunication networks.

\section*{Acknowledgements}
\label{section-Acknowledgements}

This work is financially supported by the Research Federation LUMAT and the C'Nano \^Ile-de-France GENEPHY project, and Margaux Barbier acknowledges the DGA for its financial support.

\appendix

\section*{Appendix A}
\label{section-AA}

Tables~\ref{table1} and \ref{table2} summarize the main characteristics of the two fibers involved in the experiments presented in this paper.

\newlength{\Lj}
\setlength{\Lj}{1.7cm}
\newlength{\Lk}
\setlength{\Lk}{1.3cm}
\newlength{\Ll}
\setlength{\Ll}{2.1cm}
\newlength{\Lm}
\setlength{\Lm}{2.0cm}
\newlength{\Ln}
\setlength{\Ln}{2.8cm}
\newlength{\Lo}
\setlength{\Lo}{2.2cm}

\newlength{\Ha}
\setlength{\Ha}{0.3\baselineskip}
\newlength{\Hb}
\setlength{\Hb}{0.4\baselineskip}
\newlength{\Hc}
\setlength{\Hc}{0.2\baselineskip}

\appendix
\setcounter{section}{1}

\begin{table}[h!]
  \caption{Liquid-core fiber: HC-1550-PM-01 from NKT Photonics filled with deuterated acetone.}
  \begin{center}
  \scriptsize
    \begin{tabular}{cccccc}
    \br
    \parbox{\Lj}{\centering Length \\ of the fiber\vspace{\Ha}} & \parbox{\Lk}{\centering Effective \\ area\vspace{\Ha}} & \parbox{\Ll}{\centering Linear \\ refractive index\vspace{\Ha}} & \parbox{\Lm}{\centering Zero-dispersion \\ wavelength\vspace{\Ha}} & \parbox{\Ln}{\centering Nonlinear \\ refractive index\vspace{\Ha}} & \parbox{\Lo}{\centering Nonlinear \\ coefficient\vspace{\Ha}} \\
    \hline
    \parbox{\Lj}{\centering \vspace{\Ha}1.05 m\vspace{\Ha}} & \parbox{\Lk}{\centering $\approx 30~\mu$m$^2$} & \parbox{\Ll}{\centering 1.36} & \parbox{\Lm}{\centering 896 nm} & \parbox{\Ln}{\centering $5.2 \times 10^{-20}$~m$^{2}$.W$^{-1}$} & \parbox{\Lo}{\centering 0.012~m$^{-1}$.W$^{-1}$} \\
    \br
    \br
    \multicolumn{6}{c}{\parbox{7cm}{\centering Group-velocity dispersion parameters \\ $\displaystyle{\beta_2\left(\omega\right) = \sum_{m = 0}^{8} K_m \omega^m}$\vspace{\Hc}}} \\
    \hline
    \multicolumn{6}{c}{\parbox{7cm}{\centering \vspace{\Ha}$K_0 = -6.7188826896154\times10^{-17}~\text{s}^2.\text{m}^{-1}$\\
    $K_1 = 1.7227759696642\times10^{-31}~(\text{s}^2.\text{m}^{-1}).\left(\text{rad/s}\right)^{-1}$ \\
    $K_2 = -1.6292821912020\times10^{-46}~(\text{s}^2.\text{m}^{-1}).\left(\text{rad/s}\right)^{-2}$\\
    $K_3 = 5.0231053368008\times10^{-62}~(\text{s}^2.\text{m}^{-1}).\left(\text{rad/s}\right)^{-3}$ \\
    $K_4 = 2.5279237557688\times10^{-77}~(\text{s}^2.\text{m}^{-1}).\left(\text{rad/s}\right)^{-4}$\\
    $K_5 = -2.8109811175602\times10^{-92}~(\text{s}^2.\text{m}^{-1}).\left(\text{rad/s}\right)^{-5}$\\
    $K_6 = 1.0539690566502\times10^{-107}~(\text{s}^2.\text{m}^{-1}).\left(\text{rad/s}\right)^{-6}$\\
    $K_7 = -1.8886822369153\times10^{-123}~(\text{s}^2.\text{m}^{-1}).\left(\text{rad/s}\right)^{-7}$\\
    $K_8 = 1.3576311149817\times10^{-139}~(\text{s}^2.\text{m}^{-1}).\left(\text{rad/s}\right)^{-8}$\vspace{\Hc}}} \\
    \br
    \end{tabular}
  \end{center}
  \label{table1}
\end{table}


\begin{table}[h!]
  \caption{Silica-core fiber: RTI-1605-UV from the Xlim Institute (Limoges, France) -- which has also been used for the experiments reported in \cite{JOSAB-30-2889}.}
  \begin{center}
\scriptsize
    \begin{tabular}{cccccc}
    \br
    \parbox{\Lj}{\centering Length \\ of the fiber\vspace{\Ha}} & \parbox{\Lk}{\centering Effective \\ area\vspace{\Ha}} & \parbox{\Ll}{\centering Linear \\ refractive index\vspace{\Ha}} & \parbox{\Lm}{\centering Zero-dispersion \\ wavelength\vspace{\Ha}} & \parbox{\Ln}{\centering Nonlinear \\ refractive index\vspace{\Ha}} & \parbox{\Lo}{\centering Nonlinear \\ coefficient\vspace{\Ha}} \\
    \hline
    \parbox{\Lj}{\centering \vspace{\Ha} 0.97 m} & \parbox{\Lk}{\centering \vspace{\Ha}$\approx 4~\mu$m$^2$} & \parbox{\Ll}{\centering \vspace{\Ha}1.45} & \parbox{\Lm}{\centering \vspace{\Ha}834 nm} & \parbox{\Ln}{\centering \vspace{\Ha}$2.7 \times 10^{-20}$~m$^{2}$.W$^{-1}$} & \parbox{\Lo}{\centering \vspace{\Ha}0.05~m$^{-1}$.W$^{-1}$} \\  
    \br
    \end{tabular}
  \end{center}
  \label{table2}
\end{table}

\section*{Appendix B}
\label{section-AB}

In this appendix, we give an overview of the technical details related to the coincident count measurements. Such measurements rely on a Time-Correlation Single-Photon Counting (TCSPC) device, which permits to measure the delays between a detection on one SPD and a detection on the other one.

In our experiments, a high-resolution histogram (with a bin width of 50~ps) showed that the electronic jitter of the TCSPC device was equal to 2.5~ns (much higher than the pump pulse duration, but lower than the 12.5-ns repetition period of the pump laser). Then, we chose to work with a bin width equal to this electronic jitter (see figure~\ref{FigureHisto}). In this way, the 2.5-ns bin that is located around the zero delay represents the \emph{coincidence window}: any measurement point (or \emph{occurrence}) recorded in this bin can be attributed to two simultaneous detections (one on each SPD). Among these coincident counts, we can distinguish the \textquotedblleft true\textquotedblright~coincident counts, due to correlated photons (i.e. the signal and idler photons of the same pair), and the \textquotedblleft accidental\textquotedblright~coincident counts, due to simultaneous counts on the two SPDs that \emph{do not} come from two correlated photons (for example, a signal photon on the first SPD and a residual pump photon on the second one). Since the dark count and environmental noise levels are very low, most of the bins surrounding the coincidence window are almost empty (see insert in figure~\ref{FigureHisto}): only those that are separated from the coincidence window by a multiple of 12.5~ns (repetition period of the pump laser) are populated. These secondary bins mainly correspond to the detection of photons associated with two different pump pulses. As these two counts are necessarily uncorrelated, it is usual to consider that \emph{the number of occurrences in a secondary bin is equal to the number of accidental coincidences in the coincidence window}. Thus, it is possible to evaluate the ratio between total and accidental coincidences (named \emph{coincident-to-accidental ratio}, CAR), which is a quantitative criterion to estimate the quantum purity of the source considering the performance of the detection setup (essentially its collection efficiency).

\section*{References}

\bibliographystyle{iopart-num}
\bibliography{BiblioTOT}


\end{document}